\documentclass[11pt]{article}

\usepackage{amsmath, amsthm, amssymb, graphicx}

\newtheorem{definition}{Definition}

\newtheorem{thm}{Theorem}[section]

\newtheorem{cor}[thm]{Corollary}

\title{An Optimal Bloom Filter Replacement Based on Matrix Solving}

\author{Ely Porat\\Bar-Ilan University}

\begin{document}
\date{}

\maketitle

\begin{abstract}
We suggest a method for holding a dictionary data structure, which maps keys to values, in the spirit of Bloom Filters. The space requirements of the dictionary we suggest are much smaller than those of a hashtable. 
%Rather then requiring space linear in the \emph{size of the stored data}, we only require space linear in \emph{the number of stored elements}. Our suggested data structure has a space complexity which is better than even the most advanced previously known Bloom Filters, while still maintaining the query time constant. 
We allow storing $n$ keys, each mapped to value which is a string of $k$ bits. 
Our suggested method requires $nk+o(n)$ bits space to store the dictionary, and $O(n)$ time to produce the data structure, and allows answering a membership query in $O(1)$ memory probes. The dictionary size does not depend on the \emph{size of the keys}. However, reducing the space requirements of the data structure comes at a certain cost. Our dictionary has a small probability of a one sided error. When attempting to obtain the value for a key that is stored in the dictionary we always get the correct answer. However, when testing for membership of an element that is not stored in the dictionary, we may get an incorrect answer, and when requesting the value of such an element we may get a certain random value. Our method is based on solving equations in $GF(2^k)$ and using several hash functions. 

% TODO: Whadday mean by sophisticated?
Another significant advantage of our suggested method is that we do not require using sophisticated hash functions. We only require \emph{pairwise} independent hash functions. We also suggest a data structure that requires only $nk$ bits space, has $O(n^2)$ preprocessing time, and has a $O(\log n)$ query time. However, this data structures requires a \emph{uniform} hash functions. 

% To construct the data structure, we use the hash functions and the stored elements to construct an equation set that allows answering queries. We require the equations generated for the hash function to be independent, and have certain other requirements of the hash function, so finding a good hash function takes O(ZZZ) time. However, this process need only be run once, as a preprocessing, and after it is done, we can quickly answer queries. 

In order replace a Bloom Filter of $n$ elements with an error proability of $2^{-k}$, we require $nk+o(n)$ memory bits, $O(1)$ query time, $O(n)$ preprocessing time, and only pairwise independent hash function. Even the most advanced previously known Bloom Filter would require $nk+O(n)$ space, and a uniform hash functions, so our method is significantly less space consuming especially when $k$ is small. 

Our suggested dictionary can replace Bloom Filters, and has many applications. A few application examples are dictionaries for storing bad passwords, differential files in databases, Internet caching and distributed storage systems.

%Our dictionary also has certain advantages over traditional Bloom Filters:
\end{abstract}

\section{Introduction}

A \emph{Bloom Filter} is a very basic data structure which, given a set of n elements, allows us to quickly decide whether a given element is in the set or not. The main advantage of Bloom Filters is that they are very memory efficient --- a Bloom Filter only requires space linear in the number of elements in the set, while other data structures use memory linear in the size of the represented elements in the set. When the elements stored in the set do not have a succinct representation, this is a very significant advantage. For example, consider strings, with average size of 800 bits. A hashtable for storing 100,000,000 such strings would require at least 800*100,000,000 bits, so a hard disk must be used for the table, and lookups would be rather slow. A basic Bloom Filter based structure would only require 145,000,000 bits, which can easily be stored in the main memory. On the other hand, the Bloom Filter achieves this at a certain cost. A Bloom Filter has a certain probability of returning a wrong answer. The error is one sided: if the key is in the set, the Bloom Filter will always return the correct answer, but if the key is not in the set, it might return a wrong answer. However, for many applications, it is possible to overcome this problem, and still gain from the low space requirements of the Bloom Filter. 

The main use of the Bloom Filter is to reduce the memory that the data structure uses. The basic Bloom Filter~\cite{bloom} (invented in 1970) used $n\log e$ memory bits and returned the answer using a single probe to memory, with error probability of $\frac{1}{2}$ (for a false positive).  One way to reduce the error probability is to run the basic Bloom Filter $k$ times, therefore it would require $nk\log e$  memory bits and $k$ memory probes in order to answer a query.  

During the past few years, several papers have been published on Bloom Filter~\cite{BM99,Mitzenmacher02,CM03,CKRT04,pagh05}. Most of which provided methods for reducing the memory and the number of probes required, but only considered the case where $k$ is big enough. One more disadvantage of these newer methods is that they do not allow ``insertion'' operations, which were possible to perform using the original Bloom Filter technique. Yet another disadvantage of these newer methods is that they require universal hash functions. Such functions are computationally inefficient, or have large memory requirements. 

In this paper we provide a new data structure that can replace Bloom Filters, and has lower space requirements. Our data structure requires $nk+o(n)$ memory bits (which is optimal up to $o(n)$), and each query takes $O(1)$ memory probes. However, like most of the other Bloom Filter replacements, our data structure is static and does not support insertions. Building our data structure requires $O(n)$ preprocessing time and $O(n)$ memory. This data structure is based on solving equations, and uses hash functions. We only require hash functions that are \emph{pairwise} independent. 

In addition, we suggest a similar data structure that requires only $nk$ memory bits, $O(\log n)$ query time, and $O(n^2)$ preprocessing time. However, this data structure requires uniform hash functions. 

\subsection{Applications of Bloom Filters}

Bloom Filters, as well as Bloom Filter replacements such as the one we suggest, have many applications. A good survey of Bloom Filter uses can be found in~\cite{broder02network}. A few examples are given below. 

\emph{Dictionaries:}
Early versions of UNIX's spell checker used a Bloom Filter of the dictionary instead of the dictionary itself. This Bloom Filter left several words misspelled, but the memory in these days was valuable resource and the memory it save was worth it~\cite{McIlroy82, MM90}.

The Bloom Filter was proposed as a method to succinctly store a dictionary of unsuitable passwords for security purposes by Spafford~\cite{Spafford92}. Manber and Wu describe a simple way to extend the technique so that passwords that are within edit distance 1 of the dictionary word are also not allowed~\cite{MW94}. In this setting, a false positive could force a user to avoid a password even if it is not really in the set of unsuitable passwords.
	
\emph{Databases:}
Bloom Filters can also be used for differential files~\cite{Gremillion82, Mullin83}.
Suppose that all the changes to a database that occur during the day are stored in a differential file and are updated back to the database only at the end of a day. During that day, every read from the database should first be checked in that differential file to be sure that the record read is the most recent. This file might be large, so reading through it can be slow, as opposed to querying a database, but still obligated. A possible solution to this problem is keeping a Bloom Filter of the records that have changed. Here, a false positive forces a read of the differential file even when a record has not been changed.

\emph{Internet Cache Protocol:}
Fan, Cao, Almeida, and Broder describe Summary Cache, which uses Bloom Filters for Web cache sharing~\cite{FCAB00}. In this setup, proxies cooperate in the following way: on a cache miss, a proxy attempts to determine if another proxy cache holds the desired Web page; if so, a request is made to that proxy rather than trying to obtain that page from the Web. For such a scheme to be effective, proxies must know the contents of other proxy caches. In Summary Cache, to reduce message traffic, proxies do not transfer URL lists corresponding to the exact contents of their caches, but instead periodically broadcast Bloom Filters that represent the contents of their cache. If a proxy wishes to determine if another proxy has a page in its cache, it checks the appropriate Bloom Filter. In the case of a false positive, a proxy may request a page from another proxy, only to find that that proxy does not actually have that page cached. In that case, some additional delay has been incurred. But the load on the proxy servers was reduced therefore making them work faster. 

\emph{Caching for Google's BigTables:}
BigTable is a distributed storage system for managing structured data that is designed to scale to a very large size: petabytes of data across thousands of commodity servers. Many projects at Google store data in BigTables, including web indexing, Google Earth, and Google Finance. These applications place very different demands on the BigTable, both in terms of data size (from URLs to web pages to satellite imagery) and latency requirements (from back end bulk processing to real-time data serving). Despite these varied demands, BigTable has successfully provided a flexible, high-performance solution for all of the above Google products. In some of the BigTable applications most of the queries aren't in the table. In BigTables Bloom Filter is used to determine whether a query is in the BigTable in first place, thus reducing disk accesses. A Bloom Filter can be also used in the client side as well to reduce the communication and latency.

\section{Outline}
The structure of this paper is as follows: In section~\ref{l_dictionary} we define the dictionary data structure and give a high-level view of our method, as well as a basic result. In section~\ref{l_improved_dictionary} we show how to improve the data structure to support queries in $O(1)$ time, and how to do the preprocessing in $O(n)$ time. In section~\ref{l_practicale_improvements} we show several methods to reduce constants hidden in these space complexity, which may be important in practice. In section~\ref{l_simple_hash_functions} we explain why and how simple pairwise independents hash function are enough. In section~\ref{l_membership_query} we show how to use the dictionary data structure in order to get a good Bloom Filter replacement.

\section{Dictionary Based on Matrix Solving}
\label{l_dictionary}

Dictionaries are data structures that hold key-value pairs. This section describes a method for concise representation of dictionaries with one sided errors, in the spirit of Bloom Filters. 

\begin{definition}
{A one sided error dictionary (U,k,n)} is a data structure that holds values for keys. It is a mapping from $x_1,x_2,\ldots,x_n\in U$ to $d_1,d_2,\ldots,d_n\in \{0,1,\ldots ,2^k-1\}$. Given a key $x_i$, a dictionary allows retrieving $d_i$. However, given a key $x$ which is not one of the $x_i$'s it may return  any value. 
\end{definition}

We now show how to build a dictionary which requires a storage space of $nk+o(n)$ bits. The high level concept behind our method is solving equations. Assume we have a fully random hash function $h$ from $U$ to $n$ variable equation in $GF(2^k)$ (we later show how to remove the fully random assumption later), i.e. $h:U\rightarrow GF(2^k)^{n}$. We go over all the $x_i$'s and we write the equation $h(x_i)\cdot \vec{b}=d_i$. We get $n$ equations with $n$ variables. If these equations are \emph{independent} we can solve them in $O(n^3)$ time. This can be done in a one time preprocessing, after which we can store the hash function $h$ and the vector $\vec{b}$ as our data structure. The vector $\vec{b}$ requires $nk$ bits space. To answer a query $x$ we apply $h$ on $x$ and compute $h(x)\cdot\vec{b}$ and return the answer. If $x$ is one of the $x_i$'s we get the correct $d_i$. If $x$ is not one of the $x_i$'s we might return an erroneous answer. The overall query time is $O(n)$.

However, this process only works when we get an \emph{independent} set of equations. We now examine the probability of obtaining such an independent equation set. 

\begin{thm}
The probability that our method generates an \emph{independent} set of $n$ equations on $n+c$ variables in the field $GF(2^k)$ is at least $1-\frac{1}{2^{kc}(2^k-1)}$
\end{thm}
\begin{proof} 

We order the generated equations according to the order in which they are constructed. The set of the equations is dependent when there exist $i$ such that equations $1,2,\ldots,i-1$ and equation $i$ are dependent. The probability that equation $i$ and equations $1,2,\ldots,i-1$ are dependent is at most $\frac{(2^k)^i}{(2^k)^{n+c}}$ (the probability is even lower when there are dependent equations before that index).
We apply the union bound and get that the probability that there exists an $i$ such that the equation $i$ and the equations before it are dependent is at most $\sum^{n}_{i=0} \frac{(2^k)^i}{(2^k)^{n+c}}<\frac{1}{2^{kc}(2^k-1)}$
\end{proof}
 
\begin{cor}
Even for $c=0$ we get an independent set of equations with constant probability. Therefore we need to run the preprocessing algorithm $O(1)$ time, each time with a different hash function, in order to get an independent set of equations.
\end{cor}

The main disadvantage of this data structure is that it requires $O(n)$ time in order to answer a query. One possible improvement can be achieved by using {$t$-sparse equations}. 

\begin{definition} {$t$-sparse equations} are equations of the form $\sum^{n}_{i=1} a_i$, where $|\{a_i|a_i\ne 0\}|\le t$. 
\end{definition}

Using $t$-sparse equations the query time shrinks to $t$ memory probes, $O(t)$ time.
However we need at least $m=n(1+e^{-t-\epsilon})$ variables in our equations set in order to have a full independent equations set.
\begin{thm}
If we have $n$ $t$-sparse random equations in less than $m=n(1+e^{-t-\epsilon})$ variables, the equations will be dependent with high probability. 
\end{thm}
\begin{proof}
When we have $n$ $t$-sparse random equations on $m=n(1+e^{-t-\epsilon})$ there are some variables that we do not use. Because we can look on it as throwing $t\times n$ balls to $m$ cells. The expected number of empty cells is $m(1-\frac{1}{m})^{tn}\approx me^{-\frac{tn}{m}}$. Therefore the expected number of variables we use in our equations is $m(1-e^{-\frac{tn}{m}})$. If $m(1-e^{-\frac{tn}{m}})<n$, we get $n$ equations on less then $n$ variables and therefore they will not be independent.
\end{proof}

Actually if we take $n(1+e^{-t})$ we will have a good probability to get independent set of equations.

Note that the preprocessing of the ``sparse'' data structure is $O(tn^2)$, using the Wiedemann algorithm~\cite{Wiedemann86} for solving sparse linear equations.

\section{Improved Dictionary}
\label{l_improved_dictionary}  

We now show how to reduce the query time to $O(1)$ memory probes. We also reduce the preprocessing time to $O(n)$. The high level idea behind the method suggested in this section is to divide $x_1,x_2,\ldots,x_n$ randomly to small buckets, and to run the same algorithm on each of the buckets. 

We can randomly hash the keys to $\frac{n}{s}$ buckets using hash function $h_1:U\rightarrow \{1,2,\ldots,\frac{n}{s}\}$. The expected number of keys in each bucket would be $s$, and if $s$ is big enough, with high probability there will not be a bucket with more then $2s$ keys (if there such a bucket we can choose another hash function $h_1$ and so on). Querying for $x$ is done by simply applying $h_1(x)$ and going to the $h_1(x)$'th data structure. In that data structure we query for $x$ as done in section~\ref{l_dictionary}. The $h_1(x)$ data structure does not contain more then $2s$ keys, so it would take $O(s)$ time to answer the query. The preprocessing is now performed by choosing $h_1$ and checking if there is no bucket with more than $2s$ keys. If there is such a bucket, we choose another hash function $h_1$. This is done $O(1)$ times. We then divide the keys $x_1,x_2,\ldots,x_n$ to the buckets and run the same preprocessing method described in section~\ref{l_dictionary} on each bucket. 

%%%@@@ 
Overall it would take $O(\frac{n}{s}s^3)=O(ns^2)$. The memory that this data structure consumes is $nk+O(\frac{n}{s}\log n)$ memory bits. The $O(\frac{n}{s}\log n)$ is required in order to maintain pointers to each of the data structures. Naturally, our method works best when $s$ is small. However, if we reduce $s$ too much we we lose the fact that with high probability there is no bucket which is bigger then $2s$, and the $O(\frac{n}{s}\log n)$ becomes significant.

We solve this problem by using a two-level hashing. We first explain the preprocessing and then show how to run a query. Given $x_1,x_2,\ldots,x_n$ we hash them using $h_1:U\rightarrow \{1,2,\ldots,\frac{n}{\log^2 n}\}$, which we now only require to be pairwise independent, to $\frac{n}{\log^2 n}$ buckets. It might be the case that there are some buckets which more then $2\log^4 n$ keys. We call such big buckets \emph{bad buckets}. We choose another $h_1$ hash function only if we will get more then $\frac{n}{\log^2n}$ keys hashed to bad buckets.

%%%%

\begin{thm}
The probability that there are more then $\frac{n}{\log^2n}$ keys hashed to bad buckets is at most $\frac{1}{2}$
\end{thm}       
\begin{proof}
We denote by $B_i$ the number of keys hashed to bucket $i$. Using Markov's inequality we get:
$$Pr\left[Bucket\ i\ is \ bad \right]=Pr\left[B_i>2\log^4 n\right]=Pr\left[B_i>2\log^2 E(B_i)\right]<\frac{1}{2\log^2 n}$$
Denote by $X_i$ the event that $x_i$ is hashed to a bad bucket, and by $X=\sum^{n}_{i=1} X_i$ the number of keys hashed to a bad bucket. $Pr\left[ X_i=1 \right]<\frac{1}{2\log^2 n}$ therefore $E(X)<\frac{n}{2\log^2 n}$. $Pr\left[ X > \frac{n}{\log^2 n} \right]<Pr\left[ X > 2E(X) \right]<\frac{1}{2}$ by markov inequality.
\end{proof}

\begin{cor}
It takes $O(n)$ time to find a hash function $h_1$ that we can use for the rest of the procedure.
\end{cor}

After we find a good hash function $h_1$, we deal with all the keys that are hashed to a bad bucket using a regular dictionary data structure. It takes at most $O(\frac{n}{\log n})=o(n)$ bits (we can easily modify it to take $O(\frac{n}{\log^c n})$ bits for any constant $c$).

Denote by $B_i$ the number of keys hashed by $h_1$ to bucket $i$. Each good bucket $i$ (such that $B_i<2\log^4 n$) is splitted again to sub-buckets using $h_{2,i}:U\rightarrow \{1,2,\ldots,\frac{B_i}{\frac{1}{2}\sqrt{\frac{\log n}{k}}}\}$ (we now assume that $h_{2,i}$ is fully random, in section~\ref{l_simple_hash_functions} we show how to relax this assumption). If we get a sub-bucket which is bigger then $\sqrt{\frac{\log n}{2k}}$ we choose another $h_{2,i}$.

\begin{thm}
When we split a bucket to a sub-buckets the probability that there exist sub-bucket which more then $\sqrt{\frac{\log n}{2k}}$ keys hashed to it is at most $\frac{1}{2}$ 
\end{thm}
\begin{proof}
The expected number of keys hashed to a sub-bucket is $\frac{1}{2}\sqrt{\frac{\log n}{k}}$.
Using Chernoff's inequality we get that the probability for each sub-bucket to have more then $\sqrt{\frac{\log n}{2k}}$ is much smaller then $\frac{1}{\log^4 n}$. Using the union bound we get that the probability that there exist a sub-bucket with more then  $\sqrt{\frac{\log n}{2k}}$ is smaller then $\frac{1}{2}$, since we have less then $\frac{\log^4 n}{2}$ sub-buckets. 
\end{proof}

\begin{cor}
It takes $O(B_i)$ time to find such an $h_{2,i}$. Overall, finding a hash function $h_{2,i}$ for all $i$'s requires $O(n)$ time. 
\end{cor}

We now have many smaller dictionary sub-problems. Each one of them has a size of less then $\sqrt{\frac{\log n}{2k}}$. We solve each one of them using the method mentioned in section~\ref{l_dictionary}. For each sub problem we get a random matrix of size bounded by $\sqrt{\frac{\log n}{2k}}\times\sqrt{\frac{\log n}{2k}}$ over $GF(2^k)$. The number of different such matrices is at most $2^{k(\sqrt{\frac{\log n}{2k}})^2}=\sqrt{n}$. Thus we can list all the different matrices and solve them in advance in time $O(\sqrt{n}\log^{1.5} n)$, and the list would require $O(\sqrt{n}\log n)$ memory bits.

Thus the preprocessing takes $O(n)$ time, since we can solve each sub-problem by simply looking in the list. 

We store the data structure as follows. We store all the keys which map to bad buckets using a regular dictionary, with $o(n)$ memory bits. We store a big array of less then $n$ words, each consisting of $k$ bits which are the concatenation of all the sub-buckets in all the buckets. We also store a select data structure which gives us the ability to jump in $O(1)$ memory probes to each of the buckets and sub-buckets. It requires $o(n)$ memory bits as well. Finally, we store all the hash functions. In section~\ref{l_simple_hash_functions} we show how they can be stored. Overall we use $nk+o(n)$ memory bits.

To answer a query we simply use $h_1$ in order to see to which of the bucket we need to go. If it is a bad bucket, we look for the query in the regular dictionary data structure. Otherwise we use $h_{2,i}$ in order to find in which sub-bucket the query falls. All the operation up to this point take $O(1)$ time, and we use one probe to the memory to retrieve $h_{2,i}$. We use the dictionary data structure of the sub-bucket in order to answer the query. It takes $O(1)$ probes to the memory (we retrieve $\sqrt{\frac{\log n k}{2}}$ bits in these probes, and in the last probe we take a word), but it takes $O(\sqrt{\frac{\log n }{2k}})$ time to retrieve the answer. In order to reduce that time to $O(1)$ we have two options: we can either use sparse equations or we can construct a table holding all the answers to all of the possible equations on all of the possibles assignments, and answer the query in $O(1)$ time by probing a table for getting the answer\footnote{We can play a little more with the size of each sub-bucket in order to do this in $o(n)$ space}.  

%%%@@@@

\section{Practical Improvements}
\label{l_practicale_improvements}

We now examine a few practical improvements for our method. 

\emph{Sparse equations: } Whenever we use the solution of section~\ref{l_dictionary} (even inside the sub-buckets) we can use $\ln n $ sparse equations set (in the sub-bucket case it is $\ln \log n$). This still works fine even when we use only $n$ variables, therefore it requires $nk+o(n)$ memory bits. Note that this will not work if we take only even number of variables per equation.

Another sparse equations improvement is to create equations which will be more or less local i.e. the $\{i |a_i\ne 0 \}$  will be close to each other. This way need less memory probes, because in each memory probe we can get $O(\log n)$ continues bits. 

\emph{Another counting argument: } If we make each sub-bucket bigger, we can gain in the $o(n)$ overhead. Denote by $s$ the maximum number of keys hashed to a sub-bucket. For each such sub-bucket (from section~\ref{l_improved_dictionary}). In section~\ref{l_improved_dictionary} we had a certain preprocessing analysis. We now give an alternative one. In each sub-bucket we hash keys to $\{1,2,\ldots,s^2\}$. With probability of at least $\frac{1}{2}$ there will not be any collision in this hash. If we do have a collision we choose another hash function. On average, $2$ bits are required to store which hash function we use in each sub-bucket. We now have a list of at most $s$ keys from the universe $\{1,2,\ldots,s^2\}$,  where each key gets a value in $GF(2^k)$. Note that if we have the same set of keys in two different sub-buckets, we can use the same set of equations even if they do not get the same values --- being a full rank equations set does not depand on the values (the free vector).
Thus, the number of different sets of equations we use is $s^2 \choose s$. For $s<\frac{\log n}{2\log\log n}$ we get $o(\sqrt{n})$ different equations sets. 
For each of the equations sets we compute the inverse and store it in a hashtable. The naive way to perform the preprocessing using this technique takes $\sum^{\#sub-bucket}_{i=1}O(sub-bucket-size^2)=O(n\frac{\log n}{\log\log n})$ time, because we need to multiply the inverse matrix by the data for each sub-bucket. However we can collect $O(\log n)$ sub-buckets that map to the same matrix (inverse matrix) and multiply the same matrix by $O(\log n)$ different values vectors. We get $O(\log n)$ speed up in time using word operations. Therefore the preprocessing running time shrinks back to $O(n)$. Making the equations $O(\ln \log n)$ sparse and local we get $O(1)$ query time as well\footnote{using tables as well}.

\emph{A real $nk$ solution: } We can get rid of the extra $o(n)$, by solving $n$ equations in $n$ variables. Each equation will be $\ln n$ sparse equation. The preprocessing time takes $O(n^2)$ using the block Wiedemann algorithm~\cite{villard-study}, and the query takes $O(\log n)$ time. Note that we need to use a uniform hash function for this result.

\section{Using simple hash functions}
\label{l_simple_hash_functions}
We only assume a truly random hash function inside the buckets. Each bucket consist of at most $\log^4 n$ keys. Therefore we can construct hash function by simply using array $R$ of $\log^8 n$ random numbers and a pairwise independent hash function $h:U\rightarrow \{0,1,\ldots,\log^8 n\}$. The result for the new hash function is $R[h(x)]$. Given that we hash at most $\log^4 n$ keys. The probability that there exist two keys that use the same random number is less then $\frac{1}{2}$. Therefore we got a random enough hash function with probability $\frac{1}{2}$. If we store $2\log $ hash functions like this, with probability bigger then $1-\frac{1}{n}$ each bucket will have at least one hash function which will satisfied it. The only extra space required is $O(\log^9 n)$ memory bits.

\section{Membership Queries}
\label{l_membership_query}

We first define a membership data structure. 

\begin{definition}
{A Membership data structure(n,k) for $x_1,x_,\ldots,x_n\in U$} is a data structure that allows answering membership queries. Given a query $x$ where $x$ is one of the $x_i$'s, the data structure always returns $1$, and given a query $x$ where $x$ is not one of the $x_i$ it returns 0 with probability of at least $2^{-k}$.  
\end{definition}

We can easily build a membership data structure given a dictionary data structure. We simply choose random pairwise independent hash function $h:U\rightarrow \{0,1,\ldots,2^k-1\}$ and we store a dictionary that map $x_i$ to $h(x_i)$. 

In order to check if $x$ is in the data structure we simply query $x$ from the dictionary data structure and check if it's value equal to $h(x)$. If $x$ is in the data structure it will always return $1$.

\begin{thm}
If $x$ isn't in the data structure we will return $1$ with probability $2^-k$.
\end{thm}
\begin{proof}
We choose the hash function independent from the dictionary data structure. Therefore the answer of the query $x$ from the dictionary data structure, if $x$ isn't a member is a $k$-bit string which is independent to $h(x)$. Then the probability  that they are equal is $2^{-k}$ because $h(x)$ is random.
\end{proof}

\section{Conclusions and Open problems}

We have suggested a new data structure that can replace Bloom Filters. This data structure allows maintaining a dictionary mapping keys to values, and allows retrieving the value for a key with a one sided error. Our method has significant advantages over Bloom Filter and other previously know Bloom Filter replacements. It uses only $nk+o(n)$ memory bits (which is optimal up to $o(n)$), and each query takes $O(1)$ memory probes. Also, we only require pairwise independent hash function. 

We have also suggested a similar data structure, that has an even lower space requirement, of only $nk$ memory bits. However, it has a $O(\log n)$ query time and requires $O(n^2)$ preprocessing time. Also, this data structure requires uniform hash functions. 

Despite its advantages, the method we suggest, like several other Bloom Filter replacements, does not allow ``insertion'' operations, which the original Bloom Filter technique does support. 

We believe the preprocessing phase of our algorithm can be distributed easily. In fact, we believe it should be distributed in most applications, due to the memory it consumes. 

There are several directions open for future research. First, it will be interesting to see if it is possible to design a data structure which only requires one pass on the input elements and with small additional memory. Also, it may be possible to develope a fully \emph{dynamic} data structure, with space requirements lower than those of the traditional Bloom Filter.

%\addcontentsline{toc}{section}{Bibliography}
\bibliographystyle{plain}
\bibliography{paper}
%\begin{thebibliography}{5}
%put citation to: dipmat.math.unipa.it/~mignosi/Download/preprint244.ps
%put k-mismatch in citation.
%\bibitem{latex}Helmut Kopka and Patrick W. Daly, \textsl{A Guide to
%\LaTeX: Document Preparation for Beginners and Advanced Users},
%third edition, Addison-Wesley (1999).

%\end{thebibliography}

\end{document}